\documentclass[aps, ,prb,longbibliography,floatfix,superscriptaddress,twocolumn,showpacs,amsmath,amssymb,reprint,nofootinbib]{revtex4-2}

\usepackage{graphicx}
\usepackage[english]{babel}
\usepackage{bm}
\usepackage{amsmath,amsfonts}
\usepackage{amsbsy}
\usepackage[colorlinks=true,linkcolor=blue,urlcolor=blue,citecolor=blue,breaklinks=true]{hyperref}
\usepackage[utf8]{inputenc}

\usepackage{graphics}
\usepackage{subcaption}
\usepackage{xcolor}
\usepackage{bbm}
\usepackage[normalem]{ulem}

\usepackage{amssymb}
\usepackage{array}
\usepackage{amsmath}
\usepackage{graphicx}\usepackage{graphics}
\usepackage{dcolumn}
\usepackage{bm}\usepackage{varioref}
\begin{document}

\title{Giant microwave absorption in the  vortex  lattice in $s$-wave superconductors}

\author{T. Liu}
\affiliation{Department of Physics, University of Washington, Seattle, WA 98195,  USA}
\author{M. Smith}
\affiliation{Materials Science Division, Argonne National Laboratory, Lemont, Illinois 60439, USA}
\author{A. V. Andreev}
\affiliation{Department of Physics, University of Washington, Seattle, WA 98195,  USA}
\author{B. Z. Spivak}
\affiliation{Department of Physics, University of Washington, Seattle, WA 98195,  USA}

\date{\today}

\begin{abstract}
In this article we  study  microwave absorption in superconductors in the presence of a vortex lattice.
We show that in addition to the conventional absorption mechanism associated with  the vortex core motion, there is another mechanism of microwave absorption, which is caused by the time-dependence of the quasiparticle density of states outside the vortex cores. This mechanism exists even in the absence of vortex  motion and provides the dominant contribution to microwave absorption in a broad interval of physical parameters.
At low frequencies, the dissipative part of the microwave conductivity $\sigma(\omega)$  is proportional to the inelastic relaxation time, $\tau_{\mathrm{in}}$, which is typically much larger than the elastic relaxation time, $\tau_{\mathrm{el}}$. At high frequencies $\sigma(\omega)$ is proportional to the quasiparticle diffusion time across the inter-vortex distance, $\tau_{\mathrm D}$, which is still larger than $\tau_{\mathrm{el}}$.
\end{abstract}

\maketitle

\section{Introduction}

Type-II superconductors placed in a magnetic field normal to the plane, which is weaker than the upper critical field $H_{c2}$, host Abrikosov vortices whose density is set by the condition that the average flux of the magnetic field per vortex is equal to the flux quantum $\Phi_{0}=\frac{\pi \hbar c}{e}$ \cite{Abricosov}. In the absence of pinning,  the Magnus force on the vortices induced by the
transport current  through the sample causes vortex motion and
dissipation.  The corresponding dissipative conductivity has been extensively studied since the work of Bardeen and Stephen~\cite{Bardeen_Stephen, NozieresVinen,GorkovRev,LarkOvchRev,LarkinVinokur} and is described by the formula,
\begin{align}\label{eq:sigma_BS}
\sigma_{BS}  \sim & \sigma_n \frac{H_{c2}}{H},
\end{align}
where $\sigma_n$ is the normal state conductivity and $H_{c2}$ is the upper critical field. In the Bardeen-Stephen theory,
dissipation arises from the friction force caused by  vortex motion.  Equation \eqref{eq:sigma_BS} may be obtained by expressing the friction force per unit line length of the vortex as  $\bm{F} = - \eta _{BS} {\bm v}_{v}$, where
$\eta_{BS}=\Phi_{0} H_{c2}\sigma_{n}/c^{2}$
is the vortex viscosity, and ${\bm v}_{v}$ is the vortex velocity.
In the flux flow regime the latter is given by
\begin{align}\label{eq:v_drift}
  {\bm v}_{v}= &\,  c \, \frac{{\bm E} \times {\bm H}}{H^2}.
\end{align}
Equating the rate of viscous energy dissipation to Joule heat,  $\frac{H}{\Phi_0}\eta_{BS} {\bm v}_v^2 = \sigma_{BS} E^2$, one obtains Eq.~\eqref{eq:sigma_BS}.

In the presence of disorder the vortex lattice is pinned, and is capable
of supporting a dissipationless current density which is smaller than the critical current, $J_{c}$.  Therefore, Eq.~\eqref{eq:sigma_BS} is relevant only in the non-linear (flux flow) regime where the transport current significantly exceeds $J_{c}$.
On the other hand, \emph{ac}-electric fields  ${\bm E}={\bm E}_{0}\cos(\omega t)$ induce dissipation in superconductors even in the linear regime.
In this case, the microwave absorption coefficient
is controlled by the dissipative \emph{ac}-conductivity $\sigma(\omega)$. The latter can be evaluated using the relation
\begin{equation}\label{eq:heat}
T \overline{ \dot{S}} =\frac{1}{2}\sigma (\omega) {\bm E}_{0}^{2},
\end{equation}
where  $T$ is the temperature, $\dot{S}$ is the entropy production rate per unit volume, and the overline $\overline{(...)}$ indicates averaging over time.
In most  articles on microwave absorption, the  dissipative conductivity  in the linear regime is evaluated phenomenologically using the Bardeen-Stephen expression for the vortex viscosity, $\eta_{BS}$, and the vortex velocity ${\bm v}'_{v} (t) $, which is modified by the pinning forces (see, for example, \cite{GittlemanRosenblum,Bonn,Coffeey}). The corresponding result, denoted by $\sigma'_{BS}(\omega)$ below, is proportional to the elastic momentum relaxation time $\tau_{el}$ in the normal state.

In this article we describe a new mechanism  of microwave absorption in superconductors  in the mixed state. This mechanism is caused by  the spectral flow of the quasiparticle  energy levels in the presence of an \emph{ac}-electric filed, and exists even in the absence of vortex motion.
We show that  in a broad range of physical parameters the dissipative part of the conductivity caused by this mechanism can be parametrically larger than $\sigma'_{BS}(\omega)$.

 \begin{figure}
\centering
\includegraphics[width=.9\linewidth]{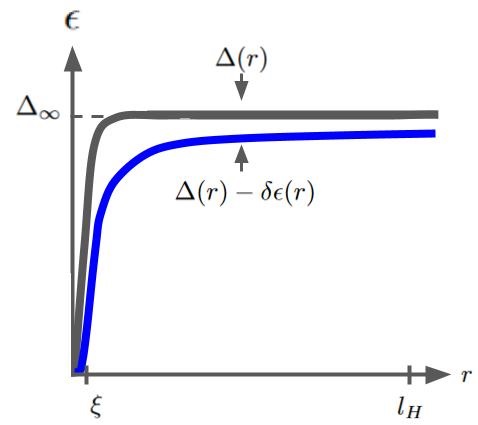}
\caption{The black line denotes the dependence of the modulus of the order parameter $\Delta(r)$ on the distance from the vortex core $r$.  The blue line shows the $r$-dependence of the edge of the quasiparticle spectrum.  }
\label{fig:FigureDOS_vortex}
\end{figure}

The origin of this mechanism can be traced to a fundamental difference between the quasiparticle kinetics in superconductors and electron kinetics in normal metals. In superconductors, the density of quasiparticle states $\nu(\epsilon,t)$   may depend on time, which means that energies of individual single particle states change in time.  In the adiabatic approximation,  the quasiparticles occupying these levels also travel in energy space.  As a result, a non-equilibrium quasiparticle distribution is created, whose  relaxation  generates entropy and provides a mechanism for microwave absorption.

The time-dependence of $\nu(\epsilon,t)$ arises because the microwave electric field induces oscillations of the superfluid momentum
\begin{align}\label{eq:p_s_def}
 \bm{p}_s(\bm{r},t)  = & \,  \frac{\hbar}{2}\left[ \bm{\nabla} \chi(\bm{r},t)     -\frac{2e}{\hbar c} \bm{A} (\bm{r},t) \right] .
\end{align}
Here $\chi(\bm{r},t) $ is the phase of the order parameter, $\bm{A} (\bm{r},t)$ - the vector potential, and $e$, $\hbar$, and $c$ are, respectively, the electron charge,  Planck's constant, and the speed of light.   The microwave field induces acceleration of the condensate,
$\dot{\bm{p}}_s (\bm{r},t) \propto e {\bm E}(t)$. Since the density of states $\nu(\epsilon,t)$ is a scalar, its time derivative can have a  linear dependence on  $\dot{\bm{p}}_{s}  (\bm{r})$  only in the  presence of a \emph{dc}-superfluid momentum $\bar{\bm{p}}_{s}({\bm r})$;   $\dot{\nu} (\bm{r})\propto  \dot{\bm{p}}_s (\bm{r}) \cdot  \bar{\bm{p}}_{s}(\bm{ r})$. In the case of a flux lattice, the \emph{dc}-superfluid momentum is the equilibrium superfluid momentum about the vortex cores.

The  non-equilibrium quasiparticle distribution created by the spectral flow can relax via two channels:  inelastic scattering  and quasiparticle diffusion. The relative importance of these  two channels depends on frequency of the microwave field $\omega$.
In typical superconductors the inelastic relaxation time, $\tau_{\mathrm{in}}$,  significantly exceeds the diffusion time across the vortex plaquette, $\tau_{\mathrm D}$. Therefore inelastic relaxation is caused by those quasiparticles for which diffusion cannot lead to full equilibration.
These quasiparticles have energies below the threshold of percolation between different vortex plaquettes.
The percolation threshold  $\epsilon^*$ arises because in the presence of a superfluid momentum $\bar{\bm{ p}}_{s} (\bm{r})$ the energy gap of the quasiparticle states is shifted down from $\Delta$ by $\delta \epsilon  (\bar{ p}_{s} (\bm{r}))$, as illustrated in Fig.~\ref{fig:FigureDOS_vortex}.
The non-equilibrium quasiparticles with  energies below  $\epsilon^*$   are trapped inside the vortex plaquettes. Because of pinning the vortex lattice  is distorted, and  the spatial distribution of the superfluid momentum $\bar{\bm{ p}}_{s} (\bm{r})$ is not symmetric about the vortex core.  Therefore, the non-equilibrium part of the of the distribution function cannot relax completely by diffusion across the plaquettes, and ultimately its  relaxation is achieved by inelastic scattering processes.
Thus, the low frequency conductivity $\sigma \sim K \tau_{\mathrm{in}}$ is proportional to the inelastic relaxation time and a parameter $K$ characterizing the degree of lattice distortion.  Since $\tau_{in}$ may exceed the elastic relaxation time $\tau_{\mathrm{el}}$ by many orders of magnitude, this contribution typically is much larger  than $\sigma'_{BS}$.
\footnote{The mentioned above mechanism of absorption has been discussed in the context of microwave absorption in superconductors in the presence of spatially uniform supercurrent \cite{Mike, Mike2} and the resistance of superconductors in flux flow regime \cite{Feigelman}. }

At higher frequencies the value of the  conductivity is controlled by the quasiparticle diffusion time across a vortex lattice plaquette, $\tau_{D}$, which is assumed to be much smaller that the inelastic relaxation time but larger than the elastic relaxation time $\tau_{\mathrm{el}} \ll \tau_{\mathrm D} \ll \tau_{\mathrm{in}}$.
In this case the value of the conductivity  $\sigma$ is still larger than $\sigma'_{BS}$.

For simplicity, we focus on the case where the thickness $d$ of the superconducting film  is smaller than the skin length, and the microwave electric field ${\bm E}(t)$  is spatially uniform  in the  film. We also assume that the distance between vortices, which is of order of the magnetic length $l_{H}=\hbar c/|e|H$, is smaller than the Pearl length~\cite{Pearl}, so that the magnetic field  is also uniform in the film.

The consideration below is organized as follows.
In Sec.~\ref{sec2} we present a general kinetic theory of quasiparticle dynamics in superconductors with time-dependent density of states.
In Sec.~\ref{sec:p_s_distribution} we discuss the spatial distribution of the condensate acceleration $\dot{\bm{p}}_s (\bm{r})$ and show that it has a significant component outside the vortex cores, which exists even in the absence of vortex motion.
 In Sec.~\ref{sec4} we use our formalism to evaluate microwave conductivity arising from this component, and show that it becomes dominant   in a broad interval of system parameters.  We discuss the results in Sec.~\ref{sec:discussion}.

\section{Quasiparticle kinetics in the presence of spectral flow}
\label{sec2}

In this section we present a general description of diffusive quasiparticle kinetics in superconductors in the presence of spectral flow and inelastic relaxation.

In  the diffusive regime, where the characteristic spatial scales exceed the elastic mean free path,
$v_{F}\tau_{\mathrm{el}}$ , the quasiparticle
distribution function $n(\epsilon, {\bm r},t)$ depends only on energy $\epsilon$, coordinate ${\bm r}$ and time $t$.
In the presence of  spectral flow, its evolution equation has the form ,
\begin{align}\label{KinEq1}
 \partial_{t} n (\epsilon, \bm{ r},  t)&+v_\nu(\epsilon,\bm{r},t)  \partial_\epsilon  n(\epsilon, \bm{r}, t) \nonumber\\
 &+ \nabla_i[ D_{ij} (\epsilon, \bm{ r},t)  \nabla_j  n (\epsilon, \bm{ r},  t)] = I_{\mathrm{in}}\{  n\}.
\end{align}
Here $ D_{ij} (\epsilon, \bm{ r},t) $ is the diffusion tensor\footnote{ Apart from the crystalline anisotropy, it acquires additional anisotropy in the presence of a superfluid momentum. In isotropic systems, it may be  expressed in terms of the longitudinal and transverse diffusion coefficients in the form $ D_{ij} (\epsilon, \bm{ r},t) = n_i n_j   D_\parallel  (\epsilon, \bm{ r},t)  + (\delta_{ij} - n_i n_j)   D_\perp  (\epsilon, \bm{ r},t) $, where $\bm{n}$ is a unit vector along $\bm{p}_s$.  It can be shown that this anisotropy can be substantial only in clean superconductors, whereas in the dirty regime it is small.}, and $I_{\mathrm{in}}\{ n\}$ is the collision integral, for which we will use the relaxation time approximation,
\begin{equation}
\label{eq:I_relaxation}
I_{\mathrm{in}}\{ n\}=- \frac{n(\epsilon)-n_{F} (\epsilon)}{\tau_{\mathrm{in}}},
\end{equation}
with $n_{F} (\epsilon)$  being the Fermi distribution function.

The second term in Eq.~\eqref{KinEq1} describes the motion of  quasiparticles in energy space caused by the spectral flow.  For sufficiently slow spatial variations of the superfluid momentum $\bm{p}_s$ and other system parameters, the ``level velocity"  $v_\nu(\epsilon, \bm{r},t)$ can be expressed in terms of the time derivative of the local  density of states, $\nu(\epsilon, \bm{r},t)$.
This relation has the same form as that in uniform systems~\cite{Mike}, and may be obtained using conservation of the number of energy levels.
From the continuity equation for the spectral current in energy space,
$\partial_t \nu(\epsilon, \bm{r}, t) +\partial_\epsilon [v_{\nu} (\epsilon,\bm{r},t)\nu (\epsilon, \bm{r},t ) ] =0,$
one gets
\begin{align}\label{eq:spectral flow}
  v_\nu(\epsilon, \bm{r},t) & = \, - \frac{1}{\nu(\epsilon,\bm{r}, t)} \int_0^\epsilon d\tilde{\epsilon} \partial_t \nu(\tilde{\epsilon},\bm{r}, t) .
\end{align}
Similarly, the diffusion coefficient  $D(\epsilon, \bm{r},t)$ in this approximation is assumed to have the same dependence on $\bm{p}_s$ and $\epsilon$ as in a uniform superconductor, $D(\epsilon, \bar{\bm p}_s(\bm r))$.

Entropy production in the system is caused by diffusion of quasiparticles as well as their inelastic relaxation. The dissipative part of the macroscopic microwave conductivity $\sigma(\omega)$ can be obtained by equating the Joule heating losses to the energy dissipation rate  Eq.~\eqref{eq:heat}.
Linearizing with respect to small deviations from the equilibrium distribution function $n=n_{F}(\epsilon)+\delta n$,
the entropy production rate is given by
\begin{align}\label{Sdot}
T\dot{S} = & \,
 T \int d \epsilon d \bm{ r}
\nu(\epsilon, \bm r, t) \bigg[ \nabla_i \delta n  D_{ij}(\epsilon, \bm r, t )   \nabla_j \delta n(\epsilon,\bm r, t) \nonumber\\
&   +
 \frac{\delta n^2(\epsilon,\bm r, t) }{\tau_{\mathrm{in}}n_{F}(\epsilon)(1-n_{F}(\epsilon))}
 \bigg].
\end{align}
Here $n_{F}(\epsilon)$ is the Fermi distribution function.

The description of quasiparticle kinetics in the presence of spectral flow given by Eqs.~\eqref{KinEq1}-\eqref{Sdot} has a broad range of applicability. It describes both clean and dirty superconductors, and does not assume a particular pairing symmetry.
On the other hand, the criteria of their applicability, and specific values of $\nu(\epsilon,\bm r, t)$,  $v_\nu(\epsilon,\bm{r},t) $, and  $D_{ij}(\epsilon, \bm r, t )$ depend on the pairing symmetry and other parameters of the system.
In the clean regime $\Delta \tau_{\mathrm{el}}\gg 1$,  Eqs.~\eqref{KinEq1} and \eqref{eq:spectral flow} have been derived in Ref.~\cite{Mike} from the conventional Boltzmann kinetic equation for quasiparticle distribution function \cite{Aronov}.
We note that Eq.~\eqref{KinEq1} has the same form as  Eq.~(A32)  in Ref.~\cite{LarkOvchRev}, which describes the time evolution of the quasiparticle distribution function $f$ in the ``dirty limit", $\Delta\tau_{\mathrm{el}}\ll 1$,
\footnote{A more general formulation of the kinetic equation for quasiparticles in dirty limit involves two distribution functions, $f$ and $f_{1}$, where the distribution function $f_{1}$ is responsible for the charge imbalance.
We neglected the latter because electromagnetic field absorption in uniform superconductors does not create  a charge imbalance.
In the weakly inhomegeneous regime considered here the generation of charge imbalance is small in the spatial gradients. At the same time,
 in the presence of pair breaking, it relaxes relatively quickly by local elastic scattering.
}
which was derived by Larkin and Ovchinnikov \cite{LarkinOvch1977}.
However, the local relation~\eqref{eq:spectral flow} between the level velocity $ v_\nu(\epsilon,\bm{r},t)$   and the time-derivative of the density of states can be derived from the  Larkin-Ovchinnikov equations only when the characteristic length scale of spatial inhomogeneity exceeds  the superconducting coherence length,  $\xi=\sqrt{D_{n}/\Delta}$,  where $D_n = v_F^2 \tau_{\mathrm{el}}/3 $, with $v_F$ being the Fermi velocity, is the normal metal diffusion coefficient.

In situations where the spectral flow is caused by the pair-breaking effect of the condensate momentum $\bm{p}_s $, the level velocity in
Eq.~\eqref{eq:spectral flow}
may be expressed in the form
\begin{equation}\label{eq:level_continuity}
v_{\nu} (\epsilon, \bm{ p}_{\mathrm{s}})=  \dot{\bm{ p}}_{s} \cdot \bm{V}(\epsilon,\bm{p}_{\mathrm s})
\end{equation}
where   $\bm V(\epsilon, \bm p_s)$  denotes the level sensitivity to changes in $\bm{p}_s$, and  is given by
\begin{equation}\label{eq:level_velocity}
   \bm{V}(\epsilon,\bm{p}_{\mathrm s}) = - \frac{1}{\nu (\epsilon,p_{\mathrm s})}  \int_{0}^{\epsilon} d \epsilon \frac{\partial \nu (\epsilon, \bm{  p}_{\mathrm s} )}{\partial \bm{p}_{\mathrm s}}.
\end{equation}
For  $s$-wave superconductors,  the  $\bm{p}_s$-dependence of the density of states and  level sensitivity was determined in Refs.~\cite{Mike,Mike2}.

Below, we use Eqs.~\eqref{KinEq1}-\eqref{eq:level_velocity} to study microwave absorption in films of type-II $s$-wave superconductors in the presence of a pinned vortex lattice.   We show that
 in the London regime, where the inter-vortex distance $\sim l_H$ exceeds the core radius $\sim \xi$, with the exception of small temperatures, $T\ll \Delta$,  the main contribution to the microwave absorption comes from quasiparticles which reside at distances $\sim l_H$  from the vortex cores.


\section{Spatial distribution of the condensate acceleration}
\label{sec:p_s_distribution}

At low frequencies  in the London regime, where the order parameter $\Delta$ outside the vortex cores is approximately uniform,
the quasiparticle density of states depends on the local instantaneous condensate momentum $\bm{p}_s(\bm{r},t) $, and its rate of change is proportional to the local condensate acceleration, $\dot{\bm{p}}_s(\bm{r},t) $.   In thin films, the Pearl length~\cite{Pearl}, which characterizes the screening of the magnetic field, practically always exceeds the inter-vortex distance, $\sim l_H $. Therefore, in this regime we can neglect the small inhomogeneity of the magnetic field ${\bm H} (\bm{r})$  caused by the diamagnetic currents.

In the presence of a microwave field ${\bm E}(t)$ the vortex positions $\bm{r}_a(t)$ become
time-dependent, and the condensate acceleration is given by,
\begin{align}
\label{eq:p_s_dot_vortices}
\dot{\bm{p}}_s (\bm{r},t) = & \,   \hbar  \hat{z} \times \sum_a \left[ \frac{  2  (\bm{r} - \bm{r}_a(t))      ( \bm{r} - \bm{r}_a(t)) \cdot \dot{\bm{r}}_a(t) }{|\bm{r} - \bm{r}_a(t)|^4}\right. \nonumber\\
&\left. - \frac{  \dot{\bm{r}}_a(t)  }{|\bm{r} - \bm{r}_a(t)|^2} \right]      + e {\bm E} (t).
\end{align}
Thus, only in the absence of vortex displacement the condensate acceleration is given by the second term, $e\bm{E} (t)$;  the modification caused by the motion of the vortices is described by the first term.

Let us consider the spatial distribution of $\dot{\bm{p}}_s (\bm{r},t)$ inside the plaquette  of a given vortex
 $a$, that is at $|\bm{r}-\bm{r}_a| \lesssim l_H$.   The term $a$ in the sum in Eq.~\eqref{eq:p_s_dot_vortices}, which corresponds to
the motion of the native core, $\dot{\bm{r}}_a$,  produces a near-field contribution which decays rapidly with the distance from the core,
\begin{align}\label{eq:near_field}
  \dot{\bm{p}}^{(n)}_s (\bm{\rho}_a  ) = & \,  \hbar \hat{z}    \left[ \frac{ 2\bm{\rho}_a\,  (\bm{\rho}_a\cdot \dot{\bm{r}}_a)   - |\bm{\rho}_a|^2 \dot{\bm{r}}_a }{|\bm{\rho}_a|^4}\right] .
\end{align}
Here we introduce the notation $\bm{\rho}_a= \bm{r}-\bm{r}_a$. The remaining contribution,  which is produced by the motion of the other vortices together with the second term, does not fall off with distance and
may be approximated by a constant for $|\bm{r} - \bm{r}_a(t)|\lesssim l_H$. We refer to it as the far field contribution and denote it by $\dot{\bm{p}}^{(f)}_s$.
By order of magnitude, $\dot{\bm{p}}^{(f)}_s$ coincides with the spatial average of the condensate acceleration in the system,  $ \langle  \dot{\bm{p}}_s (t) \rangle$.
In the linear approximation in the microwave field and at sufficiently low frequencies, where the viscous forces are negligible in comparison to pinning,  $ \langle  \dot{\bm{p}}_s (t) \rangle$ is related to the electric field by the Campbell formula,
\begin{align}\label{eq:p_s_bar_dot_E}
  \dot{\bm{p}}^{(f)}_s  \sim & \,  \langle \dot{\bm{p}}_s (t) \rangle  =  \frac{\lambda_L^2}{\lambda_{C}^2(H)} \, e {\bm E} (t).
\end{align}
Here $\lambda_L$ is the London length, and $\lambda_{C}(H)$ is the Campbell~\cite{Campbell} length. The latter  depends on the pinning strength and  characterizes the macroscopic  superfluid density of the system\footnote{   Equation \eqref{eq:p_s_bar_dot_E} reflects the fact that  the time-derivative of the superfluid transport current can be expressed  in two equivalent forms,
$\langle \dot{\bm{j}} \rangle =  \frac{c^2}{4\pi e \lambda_L^2} \langle \dot{\bm{p}}_s (t) \rangle =  \frac{c^2}{4\pi e \lambda_{C}^2 (H)} e {\bm E} (t)$.}.
Similarly, the typical vortex velocity $ \dot{\bm r}_a $, which determines the magnitude of
 $\dot{\bm{p}}^{(n)}_s (\bm{u}_a ) $ may be estimated as
\begin{align}\label{eq:v'_estimate}
  \dot{\bm{r}}_a & \sim \, {\bm v}'_v (t) =  c \,  \frac{{\bm E}(t)\times {\bm H}}{H^2} \left( 1-  \frac{\lambda_L^2}{\lambda_{C}^2(H)}   \right).
\end{align}
The second term in the brackets on the right hand side describes the pinning-induced reduction of the average vortex velocity from the value in Eq.~\eqref{eq:v_drift}.

Using the expression for the typical vortex velocity ${\bm v}'_{v} (t) $ in Eq.~\eqref{eq:v'_estimate} the Bardeen-Stephen contribution to the conductivity for a pinned vortex lattice may be estimated as
\begin{align}
\label{eq:sigma_BS'_estimate}
	\frac{\sigma_{BS}'}{\sigma_{BS}} &\sim \left(1- \frac{\lambda_L^2}{\lambda_{C}^2(H)} \right)^2.
\end{align}

Assuming that the pinning strength is determined by the vortex cores and is independent of the magnetic field, the ratio of the Campbell and London length may be expressed in the form
~\cite{Campbell,Pashinsky}
\begin{align}
    \label{eq:lambda_ratio}
   \frac{ \lambda^2_{C} (H)}{\lambda^2_L} = & \,    1+ \frac{\Phi_0 H d}{8\pi \lambda_L^2 k} .
\end{align}
Here $k$ is the average  ``spring constant", which relates the average pinning force on the vortex $\bm{F}_{pin} = - k  \delta \bm{r}$, to the average vortex displacement $\delta \bm{r}$.

The difference $ \lambda^2_{C}(H)/\lambda^2_L -1  $ characterizes the effectiveness of pinning. At perfect pinning, $k \to \infty$, we  have $ \lambda^2_{C}(H)/\lambda^2_L - 1 \to 0$ and $\langle \dot{\bm{p}}_s \rangle \to e{\bm E}$.  At finite $k$,  pinning becomes more effective as the magnetic field is reduced;   $ \lambda^2_{C}(H)/\lambda^2_L -1  $ decreases.
In particular at strong pinning, where the condensation energy of the vortex core changes by a factor of the order unity  when the vortex is displaced by a distance $\xi$ from the equilibrium position, we have   $k \sim d \Delta^2  \nu_n$.
In this case the pinning effectiveness from Eq.~\eqref{eq:lambda_ratio} may be estimated as
\begin{align}\label{eq:lambda_difference}
  \frac{\lambda^2_C(H)}{\lambda_L^2} - 1 \sim & \, \frac{H}{H_{c2}}.
\end{align}
Here we used the fact that for dirty superconductors $\lambda_L^{-2} \sim  \Delta \sigma_n/\hbar c^2  \sim  e^2 \nu_n D_n \Delta /\hbar c^2 $.

In the next section, we evaluate the dissipation arising from the two contributions to the condensate acceleration. The dissipation caused by the near field part of the condensate acceleration, $\dot{\bm{p}}^{(n)}_s (\bm{\rho}_a)$,   is localized to the vortex cores and corresponds to the Bardeen-Stephen contribution to the conductivity, Eq.~\eqref{eq:sigma_BS'_estimate}. The dissipation due to the far field part, $\dot{\bm{p}}^{(f)}_s$, arises from quasiparticles outside the vortex cores. We will show that these quasiparticles produce the dominant contribution to the conductivity in a wide interval of physical parameters.

\section{Estimates of microwave conductivity}
\label{sec4}

We now  apply   Eqs.~\eqref{KinEq1}, \eqref{eq:I_relaxation}, and \eqref{Sdot}  to evaluate the microwave conductivity $\sigma (\omega)$ in the London regime, where the inter-vortex distance $\sim l_H$ significantly exceeds the vortex core size $\sim \xi$ . We focus on  the contribution to $\sigma (\omega)$, which arises from quasiparticles residing outside the vortex cores, where the level velocity is described by Eqs.~\eqref{eq:level_continuity} and  \eqref{eq:level_velocity}. For simplicity we assume $T\sim \Delta$ and focus on the dirty limit, $\Delta \tau_{\mathrm{el}}\ll 1$.
 In this case the dependence of the quasiparticle density of states on the local superfluid momentum $\bar p_s(\bm r)$ is characterized by the dimensionless parameter
\begin{align}\label{eq:eta_def}
  \eta (p_s) &  = D_n p_s^2/\Delta.
\end{align}
The gap in the quasiparticle spectrum is lowered from $\Delta$ by the amount
\begin{align}\label{eq:delta_epsilon_estimate}
 \delta \epsilon (p_{s})  & \sim  \Delta \eta^{2/3}({p_{s} }) .
\end{align}
For  $\epsilon > \Delta + \delta\epsilon(p_s)$, the sensitivity  $\bm{V}(\epsilon, \bm p_s)$ and $\nu(\epsilon, p_s)$ are rapidly
 decreasing functions of energy $\epsilon$~\cite{Mike}. Therefore, the dominant contribution  to the microwave conductivity arises from quasiparticles with energies $ |\epsilon -\Delta | \lesssim \delta \epsilon(p_s) $.
In this energy interval the values of the level sensitivity $\bm{V}(\epsilon, \bm{p}_s  ) $, and the density of states $\nu(\epsilon, p_s)$ may be estimated as~\cite{Mike},
\begin{subequations}
 \label{eqs:PrevResults}
\begin{align}
\label{eq:DOSpsdep}
 \bm{ V}(\epsilon, \bm{p}_s  )    &\sim  D_n \bm{p}_{s} \eta^{-1/3}(p_{s} ) ,\\
\nu(\epsilon, p_s ) &\sim \nu_n \eta^{-1/3}(p_s),
\label{eq:nu_estimate}
\end{align}
\end{subequations}
where $\nu_n$ is the normal state density of states at the Fermi level.

Although in the presence of superfluid momentum the diffusion tensor is anisotropic, for $\Delta \tau_{\mathrm{el}}\ll 1$  the anisotropy is negligible,  and we set  $D_{ij}=\delta_{ij}D(\epsilon, \bar{p}_s)$. The value of the diffusion coefficient  in the relevant energy interval may be estimated as
\begin{align}
\label{eq:DiffD}
 D\left( \bar{p}_{s}  \right ) \sim   D_{n} \eta^{1/3}.
\end{align}
This estimate can be obtained\footnote{
In general the diffusion coefficient is expressed in terms of the quasiclassical Green's functions $g^R$ and $g^A$ (see the appendix in \cite{LarkOvchRev}), which must be calculated using Usadel's equation \cite{Usadel}. Since the Usadel's equation has already been solved in the relevant regime \cite{Mike2}, we will omit the details of this calculation.} using the Larkin-Ovchinnikov equations~\cite{LarkinOvch1977}.

\subsection{Low frequency regime}

At the lowest frequencies, the microwave conductivity is dominated by inelastic relaxation processes. The reason  is that diffusion cannot  lead to full relaxation of the nonequilibrium quasiparticle density for  quasiparticles with energies below the percolation threshold $\epsilon^{*}$. Such quasiparticles are trapped inside a plaquette of  a  particular vortex.  The size of the trapping region  $r(\epsilon)$ for energy $\epsilon$ may be estimated
 using Eqs.~\eqref{eq:eta_def} and  \eqref{eq:delta_epsilon_estimate} and the
the dependence of the superfluid momentum $\bar{p}_s(r)$  on the distance $r$ to  the vortex core. Since for $r \lesssim l_H$
\begin{align}\label{eq:p_s_bar_r}
 \bar{p}_s(r)\sim & \, \frac{\hbar}{r},
\end{align}
we get,
\begin{align}\label{eq:trapping_size}
  r(\epsilon) &  \sim \xi \left( \frac{\Delta}{\Delta - \epsilon}\right)^{3/4}.
\end{align}
The size of the trapping region increases with the quasiparticle energy $\epsilon$ and becomes of the order of the inter-vortex distance $\sim l_H$  as  $\epsilon$  approaches the percolation threshold $\epsilon^*$ whose value may be estimated as
\begin{align}\label{eq:percolation_threshlod}
 \Delta - \epsilon^*  & \sim \Delta \left( \frac{\xi}{l_H}\right)^{4/3}.
\end{align}
At frequencies smaller than the inverse diffusion time across the inter-vortex distance, $\tau_{\mathrm D}^{-1} $  the nonequilibrium distribution function of the trapped quasiparticles becomes spatially-uniform and depends only on the energy. This part of the distribution (zero mode of diffusion)  can relax only via inelastic processes.

To describe this slow time relaxation of the zero mode we
 linearize Eq.~\eqref{KinEq1} with respect to $\delta n$ and average the result over  the area of spatial confinement at energy $\epsilon$ in  the $a$-th  vortex. This yields
\begin{equation}\label{KinEq2}
\left( \partial_{t} + \frac{1}{\tau_{\mathrm{in}}} \right) \langle \delta n (\epsilon,t)\rangle _a  = -  \langle \dot{{\bm p}}_{s} (\bm{r})\cdot  \bm{V} (\epsilon, \bar{\bm{p}}_{\mathrm s}({\bm r}))\,\rangle  _{a}  \partial_\epsilon   n_{F} (\epsilon),
\end{equation}
where  $\langle \ldots \rangle _{a}$ denotes averaging over the $\epsilon$-dependent confinement region of the $a$-th vortex.

It is important to note that in the case of perfectly symmetric lattice the right hand side of Eq.~\eqref{KinEq2} vanishes.
However in the presence of disorder, the superfluid momentum  around the vortices is asymmetric, and this term is nonzero.
Thus, $\delta  \langle n (\epsilon)\rangle _{a}$ and  $\langle \dot{{\bm p}}_{s}(\bm{r})\cdot  \bm{V} (\epsilon, \bar{\bm{p}}_s(\bm{ r}))\,\rangle  _{a}  $
are random quantities which fluctuate from plaquette to plaquette.  Equations~\eqref{eq:DOSpsdep} and \eqref{eq:p_s_bar_r} show that these quantities are dominated by distances from the core, which are of the order of the radius of the trapping region in Eq.~\eqref{eq:trapping_size}. In this region $\dot{{\bm p}}_{s}(\bm{r})$ may be approximated by $\dot{\bm{ p}}^{(f)}_{s}(\bm{r})$  in  Eq.~\eqref{eq:p_s_bar_dot_E}.
Making this approximation,
substituting   $\delta \langle n (\epsilon)\rangle _{a}$
from Eq.~\eqref{KinEq2} into  Eq.~\eqref{Sdot}, averaging over plaquettes,
 and using Eq.~\eqref{eq:heat},  we obtain the following result for the microwave conductivity $\sigma(\omega)$
at  $T\gtrsim \Delta$,
\begin{align}
\label{eq:sigma_ratio}
\frac{\sigma (\omega) }{\sigma'_{BS}}  \sim & \,
 \frac{\tau_{\mathrm{in}}}{\tau_{\mathrm{el}}}    \frac{1}{\left[ 1 + \left( \omega \tau_{\mathrm{in}}\right)^2\right]} \, \frac{H}{H_{c2}}
\frac{\lambda_L^4(H)}{\left(\lambda_C^2(H) -  \lambda_L^2 \right)^{2}}
 \nonumber \\
 & \times
  \int_{0}^{\epsilon^{*}}  d \epsilon
  \frac{\left\langle  \langle \nu (\epsilon, \bar{p}_{s}(\bm{ r}) \rangle_{a}    \langle {\bm n}\cdot \bm{ V}(\epsilon, \bm{r}) \rangle_{a} ^{2}\right\rangle}{  T \nu_n v_{\mathrm F}^2 },
 \end{align}
 where the outer brackets indicate averaging over plaquettes, and  we introduced the unit vector ${\bm n}$  in the direction of ${\bm E}$. To arrive at this expression we used  Eqs.~\eqref{eq:sigma_BS}, \eqref{eq:sigma_BS'_estimate}, and  the Einstein relation for the normal state conductivity,
 $\sigma_{n}=e^{2} D_{n}\nu_{n}$. We  note that for the zero mode limit only the second term in the brackets  in Eq.~\eqref{Sdot} contributes to the entropy production rate.

Substituting the estimates \eqref{eqs:PrevResults}, \eqref{eq:p_s_bar_r}, and \eqref{eq:trapping_size} into Eq.~\eqref{eq:sigma_ratio} we find that the dominant contribution to microwave conductivity arises from trapped quasiparticles with energies near the percolation threshold $\epsilon^*$ given by Eq.~\eqref{eq:percolation_threshlod}. Using  \eqref{eq:lambda_ratio}, and the relation $\xi^2/l_H^2 \sim H/H_{c2}$, we can express the result in the form
\begin{equation}\label{eq:sigma_inelastic_estimate}
  \frac{\sigma (\omega)}{\sigma_{BS}'} \sim  K
   \frac{\Delta}{T}   \frac{\tau_{\mathrm{in}} \Delta}{\left[1 + \left( \omega \tau_{\mathrm{in}}\right)^2\right]}  \frac{\lambda_L^4(H)}{\left[ \lambda_C^2(H) -  \lambda_L^2 \right]^{2}} \left(\frac{H}{H_{c2}}\right)^{5/3},
\end{equation}
where the parameter
\begin{equation}\label{eq:K}
K=\frac{\left\langle     \langle {\bm n} \cdot \bm{V}(\epsilon^*, \bm{ r}) \rangle_{a} ^{2}\right\rangle }{\langle \langle  V(\epsilon^*, \bar{ p}_s(\bm{ r}) \rangle_a^2\rangle }
\end{equation}
characterizes the degree of spatial asymmetry of the condensate momentum $\bm{p}_s(\bm{r})$ inside the trapping areas in the pinned vortex lattice.
The value of $K$ is non-universal and depends on the details of the pinning potential.  If  the relative amplitude of fluctuations of $\bm{V}(\epsilon,\bm{ r})$ is of order unity, and their correlation radius is on the order of the plaquette size, then $K\approx 1$. Using the estimate \eqref{eq:lambda_difference} we obtain for strong pinning,
\begin{equation}\label{eq:sigma_inelastic_estimate_strong}
  \frac{\sigma (\omega)}{\sigma_{BS}'} \sim
   \frac{\Delta}{T}   \frac{\tau_{\mathrm{in}} \Delta}{\left[1 + \left( \omega \tau_{\mathrm{in}}\right)^2\right]}  \left(\frac{H_{c2}}{H}\right)^{1/3}.
\end{equation}
Since in typical situations   $\tau_{\mathrm{in}}\Delta \gg 1$, Eqs.~\eqref{eq:sigma_inelastic_estimate} and \eqref{eq:sigma_inelastic_estimate_strong} show that  the inelastic relaxation gives the main contribution to the conductivity  in the low frequency regime.

\subsection{High frequency regime}

At $\omega\tau_{\mathrm{in}} >1$ the contribution of inelastic relaxation to $\sigma(\omega)$ in Eq.~\eqref{eq:sigma_inelastic_estimate} decreases as $1/\omega^2$, and at sufficiently high frequencies the microwave conductivity is dominated by spatial diffusion of nonequilibrium quasiparticles inside the trapping regions.

The diffusive contribution to dissipation is present even in a perfect vortex lattice. Therefore, to estimate it we neglect the distortion of the vortex lattice. We
assume the axially symmetric distribution of the superfluid momentum  $\bar{\bm{p}}_s$ inside a given plaquette, Eq.~\eqref{eq:p_s_bar_r},
and write the diffusion equation Eq.~\eqref{KinEq1} in  polar coordinates, $r$ and $\theta$.
The dependence of the  level velocity $v_\nu (r, \theta)$  in Eq.~\eqref{eq:level_continuity} and nonequilibrim quasiparticle density $\delta n$ on the azimuth angle corresponds to the first angular harmonic,
\[
\delta n (\epsilon, r,\theta; t ) =  n_1 (\epsilon, r;t ) \cos \theta.
\]
Substituting this form into the linearized Eq.~\eqref{KinEq1} we get
\begin{align}\label{eq:kineq_polar}
&\partial_t n_1(\epsilon, r;t) + \frac{1}{r} \partial_r \left( r D(\epsilon, \bar p_s(r)) \partial_r n_1 (\epsilon, r;t)\right)\nonumber\\
& - \frac{D(\epsilon, \bar p_s(r) )n_1(\epsilon, r;t)}{r^2} = - \langle\dot p_s^{(f)}\rangle V(\epsilon, \bar p_s(r)) \partial_\epsilon n_F.
\end{align}
Here we  neglected the inelastic collision integral and focused on the long-distance part  of the condensate acceleration given by Eq.~\eqref{eq:p_s_bar_dot_E}.

For the relevant quasiparticle energies, the diffusion coefficient $ D(\epsilon, \bar p_s(r))$, and the level sensitivity $ V(\epsilon, \bar p_s(r))$ have a power-law dependence of $r$, which is given by Eqs.~\eqref{eq:eta_def}, \eqref{eq:DOSpsdep}, \eqref{eq:DiffD}, and \eqref{eq:p_s_bar_r}. 
For a microwave field of frequency $\omega$, the solution of this equation is characterized by the length scale
 $L_{\omega}$  corresponding to the diffusion distance of the relevant quasiparticles during the oscillation period,   $D\left( \bar{p}_s( L_\omega ) \right)/ L_{\omega}^{2} = \omega$,
 and may be estimated as
 \begin{align}
 	L_\omega &\sim \xi \left(\frac{\Delta}{\omega}\right)^{3/8}.
 \end{align}
For $\xi \ll r \ll L_\omega$, and $L_\omega\ll r$  the solution to Eq.~\eqref{eq:kineq_polar} has a power-law form,
\begin{align}\label{sol3}
n_1  \sim &\, e E\frac{\lambda_L^2}{\lambda^2_C(H)}\partial_\epsilon n_F
\left\{
\begin{array}{cc}
  \frac{r^{7/3}}{\xi^{4/3}}, & r \ll L_\omega, \\
  -i \frac{L_\omega^{8/3}}{r^{1/3} \xi^{4/3}},  & r \gg L_\omega.
\end{array}
\right.
%
\end{align}
Substituting Eq.~\eqref{sol3} into the first term of the RHS of  Eq.~\eqref{Sdot} and using  Eq.~\eqref{eq:p_s_bar_dot_E} we get
\begin{align}\label{eq:sigamHighFr}
\frac{\sigma (\omega)}{\sigma_{BS}'} \sim &\,  \frac{\Delta}{T} \frac{\lambda_L^4}{(\lambda_C^2-\lambda_L^2)^2} \!  \left( \frac{H}{H_{c2}}\right)^{1/3}\!\!
\left\{
\begin{array}{cc}
  1, &  \omega \ll \tau_D^{-1} , \\
  (\omega\tau_D)^{-5/4}, & \omega \gg \tau_D^{-1}.
\end{array}
\right.
%
\end{align}
Here $\tau_D$ the diffusion time of the relevant quasiparticles across the plaquette, $\tau_D = l_H^2/ D(\bar p_s(l_H))$. It may be estimated as
\begin{align}\label{eq:tau_D}
	\tau_D \sim   & \,  \frac{\hbar}{\Delta}  \left( \frac{l_H}{\xi}\right)^{8/3} .
\end{align}

Comparing Eq.~\eqref{eq:sigma_inelastic_estimate} and Eq.~\eqref{eq:sigamHighFr}  we conclude that, for  $\omega< \tau^{-1}_{\mathrm D}$, diffusive relaxation given the dominant contribution to the conductivity  for $\omega \gtrsim \omega^*$, where the crossover frequency $\omega^*$ may be estimated as
\begin{align}\label{eq:omega_star}
\omega^{*} \sim & \,  \sqrt{\frac{K \Delta}{\tau_{\mathrm{in}}}}\left (\frac{H}{H_{c2}}\right)^{2/3} \sim (\tau_{\mathrm{in}}\tau_{\mathrm D})^{-1/2}.
\end{align}

At strong pinning, see
 Eq.~\eqref{eq:lambda_difference},  the expression for the conductivity in Eq.~\eqref{eq:sigamHighFr} simplifies to
\begin{align}\label{eq:sigma_diffusive_strong}
\frac{\sigma (\omega)}{\sigma_{BS}'} \sim &\,  \frac{\Delta}{T}   \left( \frac{H_{c2}}{H}\right)^{5/3}\!\!
\left\{
\begin{array}{cc}
  1, &  \omega \ll \tau_D^{-1} , \\
  (\omega\tau_D)^{-5/4}, & \omega \gg \tau_D^{-1}.
\end{array}
\right.
\end{align}
In this case, in a broad frequency interval, $\omega^* < \omega < \tau_D^{-1}$, the \emph{ac} conductivity is dominated by diffusion of quasiparticles outside the vortex cores for sufficiently weak magnetic fields  $H \ll  H_{c2}$.

Finally, we note that replacing $\langle\dot{p}_s^{(f)} (r  ) \rangle$ in Eq.~\eqref{eq:kineq_polar} by  $\dot{p}_s^{(n)} (r  ) $ from Eq.~\eqref{eq:near_field} (which has the same angular dependence)  gives a contribution to dissipation, which is dominated by short distances from the core, and reproduces the  Bardeen-Stephen result $\sigma'_{BS}$. 

\section{discussion of the results}
\label{sec:discussion}

We developed a theory of microwave absorption of type-II superconductors in the presence of a strongly pinned vortex lattice.
In this case, in addition to the Bardeen-Stephen contribution
to the dissipative conductivity,   $\sigma'_{BS}$,
 which is caused by the vortex motion and is described by Eqs.~\eqref{eq:sigma_BS}, \eqref{eq:sigma_BS'_estimate}, and \eqref{eq:lambda_ratio}, there is an another contribution. This new contribution is caused by  the quasiparticle spectral flow induced by the condensate acceleration  in the presence of a microwave field. This contribution  exists even in the absence of vortex displacements and exceeds $\sigma_{BS}'$ in a wide interval of physical parameters.
 For  $T\gtrsim \Delta$
this contribution is dominated by the quasiparticles residing outside the vortex cores.

At low frequencies the relaxation of the non-equilibrium distribution quasiparticles induced by the spectral flow is mediated by inelastic processes. As a result, the dissipative conductivity $\sigma (\omega)$ in this regime is controlled by the inelastic relaxation time $\tau_{\mathrm{in}}$, see Eq.~\eqref{eq:sigma_inelastic_estimate}. At $\omega \ll \tau_{\mathrm{in}}$ the lowest frequencies $\sigma (\omega)$ becomes
proportional to  $\tau_{\mathrm{in}}$. Therefore it can be parametrically larger than $\sigma'_{BS}$,  which is proportional to the elastic mean free time $\tau_{\mathrm{el}}$. The mechanism  of absorption in this case is similar to the Debye mechanism of electromagnetic wave absorption is molecules \cite{Debye}. The difference however is that the Debye absorption coefficient in molecules is quadratic in frequency, while the dissipative part of conductivity in superconductors $\sigma (\omega)$, is nonzero at  $\omega\to 0$. The low frequency conductivity is proportional to the degree of distortion of the spatial distribution of the superfluid momentum in the lattice, which is caused by the pinning and characterized by the parameter $K$ in Eq.~\eqref{eq:K}.


%

At frequencies above  $\omega^*$   given by Eq.~\eqref{eq:omega_star}   the conductivity is dominated by  diffusion of quasiparticles across the vortex. The characteristic time scale $\tau_{D}$ for this process is given by   Eq.~\eqref{eq:tau_D},
and is still much larger than $\tau_{\mathrm{el}}$.  In this regime the conductivity exhibits a non-trivial frequency dependence given by Eq.~\eqref{eq:sigamHighFr}.  The distortion of the vortex lattice induced by the pining becomes inessential in this regime.


At small temperatures, $T\ll \Delta$, the  quasiparticles reside only in the vortex cores. In this case the low frequency contribution to conductivity, which is proportional to $\tau_{\mathrm{in}}$ still exists, but the coefficient $K$ in Eq.~\eqref{eq:K} is determined by the deformation of the core.

We note that in the case of weak pinning there is
another mechanism of low frequency microwave absorption, which is proportional to  $\tau_{\mathrm{in}}$, \cite{Pashinsky}. It  is related to vortex motion and exists also in the flux flow regime \cite{Feigelman}. The main contribution to $\sigma$ in this case comes from quasiparticles in the vortex cores.

Although the estimates for the micirowave conductivity  were obtained  in the dirty regime, $\Delta\tau_{\mathrm{el}}\ll 1$, qualitatively,  our conclusions remain valid for arbitrary value of $\Delta \tau_{\mathrm{el}}$, provided $l_H \gg v_F \tau_{\mathrm{el}}$. In the clean case,  $\Delta\tau_{\mathrm{el}}\gg1$ one would need to use different estimates for the sensitivities and relevant intervals, which can be found in Ref.~\cite{Mike}.

We would like to thank M. Feigel'man and V. Geshkenbein for useful discussions.
The  work of T.L. and A.A. was supported by the National Science Foundation through the MRSEC Grant
No. DMR-1719797, the Thouless Institute for Quantum
Matter (TIQM), and the College of Arts and Sciences at
the University of Washington.

\end{document}